\documentclass[letterpaper,twocolumn,10pt]{article}
\usepackage{usenix2019_v3}

\usepackage{tikz}
\usepackage{amsmath}

\usepackage{pifont}
\usepackage{booktabs}
\usepackage{amssymb}
\usepackage{enumitem}
\usepackage{titlesec}
\titlespacing*{\section}{0pt}{*1.2}{*0.3}
\titlespacing*{\subsection}{0pt}{*0.8}{*0.3}
\titlespacing*{\subsubsection}{0pt}{8pt}{4pt}

\usepackage[letterpaper, top=1in, bottom=1in, left=1in, right=1in]{geometry}
\usepackage{graphicx}    
\usepackage{subcaption}  
\usepackage[font=small]{caption}     
\usepackage{geometry}    

\usepackage{algorithm}
\usepackage{algorithmic}

\usepackage{multirow}

\usepackage{listings}
\usepackage{xcolor}

\usepackage{filecontents}

\begin{document}

\date{}

\title{\Large \bf Mycelium: A Transformation-Embedded LSM-Tree}

\author{
{\rm Holly Casaletto}\\
UC Santa Cruz
\and
{\rm Jeff Lefevre}\\
UC Santa Cruz
\and
{\rm Aldrin Montana}\\
UC Santa Cruz
\and
{\rm Peter Alvaro}\\
UC Santa Cruz
} 

\maketitle

\begin{abstract}
Compaction is a necessary, but often costly background process in write-optimized data structures like LSM-trees that reorganizes incoming data that is sequentially appended to logs. In this paper, we introduce Transformation-Embedded LSM-trees (TE-LSM), a novel approach that transparently embeds a variety of data transformations into the compaction process. While many others have sought to reduce the high cost of compaction, TE-LSMs leverage the opportunity to embed other useful work to amortize IO costs and amplification. We illustrate the use of a TE-LSM in \textit{Mycelium}, our prototype built on top of RocksDB that extends the compaction process through a cross-column-family merging mechanism. Mycelium enables seamless integration of a transformer interface and aims to better prepare data for future accesses based on access patterns. We use Mycelium to explore three types of transformations: splitting column groups, converting data formats, and index building. In addition to providing a cost model analysis, we evaluate Mycelium's write and read performance using YCSB workloads. Our results show that Mycelium incurs a 20\% write throughput overhead—significantly lower than the 35\% to 60\% overhead observed in naive approaches that perform data transformations outside of compaction—while achieving up to 425\% improvements in read latency compared to RocksDB baseline.

\end{abstract}

\section{Introduction}

The physical organization of data on storage devices dictates how efficiently it can be accessed later~\cite{athanassoulis2016designing}. However, in many modern data management systems, the exponential growth of data generation and increasingly diverse workloads leave application designers with little flexibility to craft optimal access methods at the point of ingestion~\cite{idreos2020key}. Although suboptimal access methods can hinder performance, the immediate priority often lies in handling high-volume data ingestion. As a result, users are frequently compelled to adopt write-optimized storage solutions~\cite{sarkar2022dissecting}, deferring performance tuning to a later stage. To transition data into a more read-efficient state, external processes such as ETL are often employed~\cite{kossmann2023extract, machado2019dod, zasadzinski2021trip}.

Relying on external processes to reshape data for optimized reads introduces several drawbacks: delayed availability~\cite{sabtu2024improving}, increased resource consumption for storage and data transfer~\cite{maharaj2022enhancing}, and a heightened risk of errors~\cite{souibgui2019data}. This trade-off is especially undesirable for users of real-time business intelligence applications, who require seamless integration of write-optimized and read-optimized storage ~\cite{xing2024aha, sadoghi2016store}. Log-structured merge (LSM) trees~\cite{o1996log}, widely adopted for write-heavy workloads, offer a partial solution by addressing this challenge over time: records enter a write-optimized system, but gradually find their way to a more read-optimized state. To achieve this, writes are aggressively buffered in memory at ingress and later written sequentially to block storage, providing the best possible write throughput. As records age, their storage is repeatedly coalesced, via a process called "compaction," to support future read-heavy workloads.

Nothing comes for free, however. The RUM Conjecture~\cite{athanassoulis2016designing} asserts that a storage engine cannot simultaneously achieve optimal performance for reads, updates (writes), and memory usage. The principle underlines the inherent compromises in LSM-based systems, which pay for their flexibility through increased memory consumption and write amplification: the same records are, over time, repeatedly read into memory and written back to disk, impacting both device lifetime and sustained write throughput. Over the past decade, considerable research has focused on mitigating the costs of compaction while enhancing its benefits. Efforts span a wide range of optimizations, including reducing compaction frequency~\cite{sears2012blsm}, exploring the design space of compaction strategies~\cite{sarkar2022constructing}, offloading distributed compaction jobs~\cite{ahmad2015compaction}, increasing compaction parallelism~\cite{zhang2014pipelined}, and optimizing data placement on disk~\cite{lu2017wisckey}. Most of these efforts primarily aim to reduce compaction cost and improve its efficiency.

In this paper, we explore an alternative approach--treating compaction instead of as the necessary evil as an opportunity to embed more data reorganization work that benefit future reads and scans. An approach of utilizing, rather than mitigating, compaction. Saxena et al. \cite{saxena2023real} propose leveraging compaction to transform data from row-wise storage to columnar storage, inspiring us to delve further into data reorganizations that impact the physical data designs. The core principle of LSM trees--involving copying data between storage tiers and reshaping it to suit the performance characteristics of each tier--presents an opportunity to piggyback additional transformations onto this process. Although this copying puts I/O and compute pressure on the storage server, which impacts its capacity to serve reads and writes, it creates a natural checkpoint for further optimization. Compaction, by design, serves the purpose of bounding read queries to keep lookup costs reasonable, but its potential for broader data reorganization remains underutilized.

We observe that when the storage system is I/O bottle-necked, there is an opportunity to further exploit these sunken costs to improve the overall performance of the system.  At each compaction step (below level 0), all of the records in a given file must be read from storage into memory, and then written to storage again. Any work that requires reading this data (e.g. building an index) can potentially share the scan. Any work that changes the representation of the data can potentially share the write as well, essentially performing "free" I/O.

We present Mycelium, a Transformation-Embedded-LSM-based (TE-LSM) storage system built on RocksDB. Mycelium allows users to implement compaction-time m-routines (modular transformer routines) designed to accelerate future queries; these m-routines are automatically piggybacked onto compaction as data flows through the LSM-tree. M-routines can express changes to the physical design of the data (e.g. transforming from string to binary formats) or the creation of auxiliary data structures (e.g. secondary indices). We showcase the flexibility of m-routines by using them to implement column group splitting, JSON-to-Flatbuffers conversions, and secondary indices creation. More work, such as materialized view creation, compression and decompression, ETL, and schema normalization, can be applied through m-routines as well.

Our evaluation shows that these techniques can achieve speedups of up to 425\% for various types of queries, while incurring only a 20\% write throughput overhead--significantly lower than the 35\% to 60\% penalty observed in approaches that perform data transformations externally. 

We make the following contributions in this paper:
\begin{itemize}[itemsep=0pt, topsep=0pt, left=12pt]
    \item We introduce TE-LSM, an extension of the conventional LSM-Tree design that enables the background compaction process to perform data transformations. This capability allows data to be seamlessly transformed from one format or physical layout to another while being compacted.
    \item We provide a programmable interface that defines a flexible and composable design space for implementing diverse data transformations.
    \item We propose a novel merging method for compaction, called Cross-Column-Family Tierveling, which directly integrates data transformations into the compaction process for greater efficiency.
    \item We present Mycelium, a RocksDB-based storage engine that supports four types of data transformations and their algebraic composition. Our evaluation using YCSB benchmarks demonstrates substantial improvements in read performance with minimal impact on write throughput.
\end{itemize}


\section{Background}
To understand how Mycelium integrates data transformations within compaction, it is essential to first review its two foundational concepts: LSM-tree compaction and data transformation.

\subsection{LSM-Tree Compaction}
Compaction is a critical background process in LSM-tree-based data platforms that reorganizes data to enhance read performance and optimize storage efficiency~\cite{o1996log}. 
There are several types of compaction, level-based compaction being the most commonly used~\cite{sarkar2022constructing}. The configuration parameter regarding the number of levels depends on the specific requirements of the application. While additional levels can make each compaction job more efficient by handling smaller data sizes at each step, they also increase write amplification since data may be rewritten multiple times, potentially several per level. Data are initially ingested into the skip-list in memory, then enter the LSM-tree at Level-0 through a flush operation, which moves data directly from the in-memory write buffer to disk without any changes to maximize performance. From there, background compaction jobs kick in to move the data from Level \textit{$i$}, where \textit{$i \in [0, N)$}, to Level \textit{$i+1$}, sequentially, until it reaches the final level, Level \textit{$N$}~\cite{dong2017optimizing}.

Files are added to Level-0 through flush operations and to other levels through compaction processes. At Level-0, files typically have overlapping key ranges because they are directly flushed from memory without merging. In contrast, files in levels other than Level-0 are sorted and non-overlapping due to the compaction process. Typically, after compacting from Level-i to Level-(i+1), all files in Level-i are removed. The resulting data either merges with existing files in Level-(i+1) if there are overlapping key ranges or forms new files in Level-(i+1) with new key ranges.

There are two common merging strategies for moving compacted data between levels in an LSM-tree: Tiering and Leveling. The key distinction between these approaches lies in how data is merged across levels. In leveling, data from a higher level—where Level 0 is considered the highest—is always merged into a lower level, resulting in a single sorted run at the lower level. This ensures that each level (except Level 0) contains only one sorted run. In Tiering, merging occurs entirely within the higher level when it becomes full. Once this level accumulates multiple runs, they are combined and moved down to the next lower level as a new sorted run. As a result, in Tiering-based LSM-trees, all levels except the last (the final destination for data) typically contain multiple sorted runs.

In Mycelium, we adopt a hybrid merging strategy called Tierveling. This approach uses Tiering for compactions that involve data transformations, moving data from one column family to another. For compactions that only apply the identity function (no transformations), we use leveling, moving data within the same column family.

\subsection{Data Transformation}
\vspace{-0.1cm}
Data transformation has long been a critical component of modern data-intensive applications~\cite{fernandes2023data}, with one of its most prominent roles being the foundation for data warehouse development~\cite{vassiliadis2009survey}. This process encompasses essential tasks such as data cleaning, validation, and preparation, serving as a vital stage within the well-established Extract, Transform, Load (ETL) pipeline~\cite{kherdekar2016technical, maharaj2022enhancing, souibgui2019data}. Traditionally, data transformation combines both manual and automated procedures to ensure data quality and usability~\cite{morcos2015dataxformer}.

In recent years, the growing diversity of workloads has intensified the demand for more flexible and adaptive data transformation processes to better align data with the specific requirements of target systems~\cite{chang2012workload, salamkar2020data}. These workload-optimized transformations often aim to enhance performance and efficiency by refining the physical representation of data through format or structural conversions. Notable examples include splitting row-wise input into columnar storage formats~\cite{roozkhosh2021relational, pivarski2017fast, li2020mainlining}, converting data between different formats~\cite{castillo2018optimizing, collett2002modelling}, and augmenting data with auxiliary structures such as secondary indices to accelerate query performance~\cite{bahle2002efficient, gani2016survey}.

Transforming row-wise data into columnar storage is a widely adopted strategy for optimizing transactional data for analytic queries. Data is typically generated in a row-oriented format with multiple attributes. However, as data ages, it becomes less likely to be accessed by Online Transaction Processing (OLTP) queries and more frequently queried by Online Analytical Processing (OLAP) workloads. OLAP queries often target a single column or a small subset of columns over a range of values to perform aggregations such as sums or averages. Splitting data into columnar storage significantly reduces I/O and deserialization costs by enabling more efficient access to smaller, relevant data segments.

Converting data between formats can significantly optimize storage efficiency. For instance, JSON is a human-readable text format but is inherently inefficient for storage. It lacks a schema, causing object field names to be stored with each object repeatedly, along with additional padding characters like { and ,. In contrast, binary formats such as Protobuf and FlatBuffers use schema definition files, allowing field names to be stored separately in a system catalog rather than within each object. As a result, converting data from a text-based format like JSON to a binary format like Protobuf or FlatBuffers can reduce file size and potentially lower I/O costs.

 Augmenting data with auxiliary structures like secondary indexes or materialized views can significantly improve query performance. In key-value stores, for example, the absence of a secondary index forces attribute-based searches to perform a full table scan, which becomes highly inefficient and often impractical for large datasets.

\section{TE-LSM Design}
We now introduce the design of the Transformation-embedded LSM-tree, which integrates data transformation tasks directly into the background compaction process. This design allows compaction and data transformation to be executed seamlessly as a single unified job.

A typical compaction process involves three main steps: ($1$) reading data from input SST files into memory, ($2$) filtering out the deleted and obsolete entries, and ($3$) merging the remaining data into new SST files written to disk. However, the second step is rate-limited by the expensive disk I/O operations of the first and third steps. The key idea behind embedding data transformation into compaction is to piggyback on the existing I/O operations (reads and writes) that compaction already performs. Since this approach leverages the most expensive operations while adding overhead only to the least costly in-memory step, we hypothesize that it provides significant benefits at a reasonable additional cost. In other words, by applying transformations concurrently with compaction, IO for the transformations are amortized by leveraging the IO necessary for traditional compaction. This makes them particularly effective for data-intensive workloads, as they retain the high write throughput of traditional LSM-trees (within an acceptable threshold) while enhancing read performance.

At a high level, compaction is the process of merging data and removing duplicates; then, standard compaction ($C$) can be seen as embedding a simple transformation function: the identity function (e.g. no-op). From this perspective, TE-LSMs simply allow other transformation functions to be applied, which we call ``extended compaction'' for brevity ($C^{T}$). In this paper, we support two classes of transformation functions: map (inputs and outputs are \(1-1\)) and flatmap (inputs and outputs are \(1-Many\)).


\subsection{Definitions}
We define several key concepts below that will be used in our discussion of the design.

\textbf{Column Family}: In RocksDB, a "column family" is analogous to a "table" in SQL databases. It serves as a logical grouping of key-value pairs that share similar characteristics or purposes. Each column family functions as a separate namespace within the key-value store.

\textbf{Destination Column Family}: In traditional LSM-trees, compaction operates within the same column family. However, in a transformation-embedded LSM-tree, compaction moves data between distinct column families to handle transformed outputs, referred to as destination column families. Unlike regular column families, these destination column families are system-internal and not exposed to the user interface. External applications must not write to these column families, as doing so could lead to data corruption or undefined system behavior.

\textbf{Transformer}: A Transformer is an object created to be associated with a column family as an attribute of it. All transformers follow the same interface to define transformation functions to be applied during compaction. Transformers produce transformed outputs, which are then written to destination column families. Multiple transformers can be attached to a single column family, enabling algebraic composition of transformations.

\subsection{APIs}
We outline the APIs that are supported in Mycelium below:

\begin{itemize}[left=0pt]
    \item \textbf{insert(T, k, v)}: Inserts the key-value pair \texttt{\(k/v\)} into the table \texttt{$T$}. The behavior of this API in Mycelium is identical to that in RocksDB.

    \item \textbf{read(T, k)}: Retrieves the value \(v\) associated with key \(k\) from table \(T\). In Mycelium databases with \textbf{split} transformations, the Mycelium layer handles reassembling the split pieces to reconstruct the full row. For other transformation types, this API behaves the same as in RocksDB.

    \item \textbf{read(T, k, [\texttt{\(v_i\)}])}: Reads the partial value \(v_i\) associated with key \(k\) from table \(T\). In Mycelium databases with \textbf{split} transformations, the Mycelium layer directs the query to the appropriate split internal table \(T'_i\) to retrieve the partial value. For other types of Mycelium transformations, the optional third parameter is a no-op.

    \item \textbf{read(T, [$k_1$, $k_2$])}: Performs a range scan to retrieve all key-value pairs with keys in the range \([k_1, k_2]\). In databases with \textbf{split} transformations, the Mycelium layer reassembles split pieces to reconstruct full rows. For other transformation types, this API functions identically to RocksDB.

    \item \textbf{read(T, [$k_1$, $k_2$], [\texttt{\(v_i\)}])}: Performs a range scan to retrieve partial values \([\texttt{\(v_i\)}]\) for all keys in the range \([k_1, k_2]\). In Mycelium databases with \textbf{split} transformations, the Mycelium layer directs the query to the appropriate split internal table \(T'_i\) to retrieve the partial value. For other types of Mycelium transformations, the optional third parameter is a no-op.

    \item \textbf{read(T, k, [\texttt{\(v_i\)}], ik)}: Reads from an index created with the prefix of column \(ik\) to retrieve a set of keys for further value lookup from table \(T\). This API is applicable only to Mycelium databases utilizing \textbf{augment} transformations.

    \item \textbf{read(T, [$k_1$, $k_2$], [\texttt{\(v_i\)}], ik)}: Performs a range scan of a secondary index with the prefix of column \(ik\) in the range of \([k_1, k_2]\), to retrieve a set of keys for further value lookup from table \(T\). This API is applicable only to Mycelium databases utilizing \textbf{augment} transformations.
\end{itemize}

\subsection{Cross Column Family Compaction} 


Cross-column-family compaction is at the core of enabling data transformation to be seamlessly integrated into the compaction process. Unlike regular compaction, where input and output belong to different levels of the same column family, and thus data simply moves from one level to another within the same LSM-tree, transformation-embedded compaction involves input from a source column family and output to one or more destination column families. In regular compaction, data is merged between levels (e.g., an SST file from an input level is merged with one from the output level into a new SST file). However, in cross-column-family compaction, the process more closely resembles a "flush operation"—multiple sorted runs in the Level-0 of the source column family are combined and "flushed" into the Level-0 of the destination column family.

\begin{figure}[h]
    \centering
    \includegraphics[width=\linewidth]{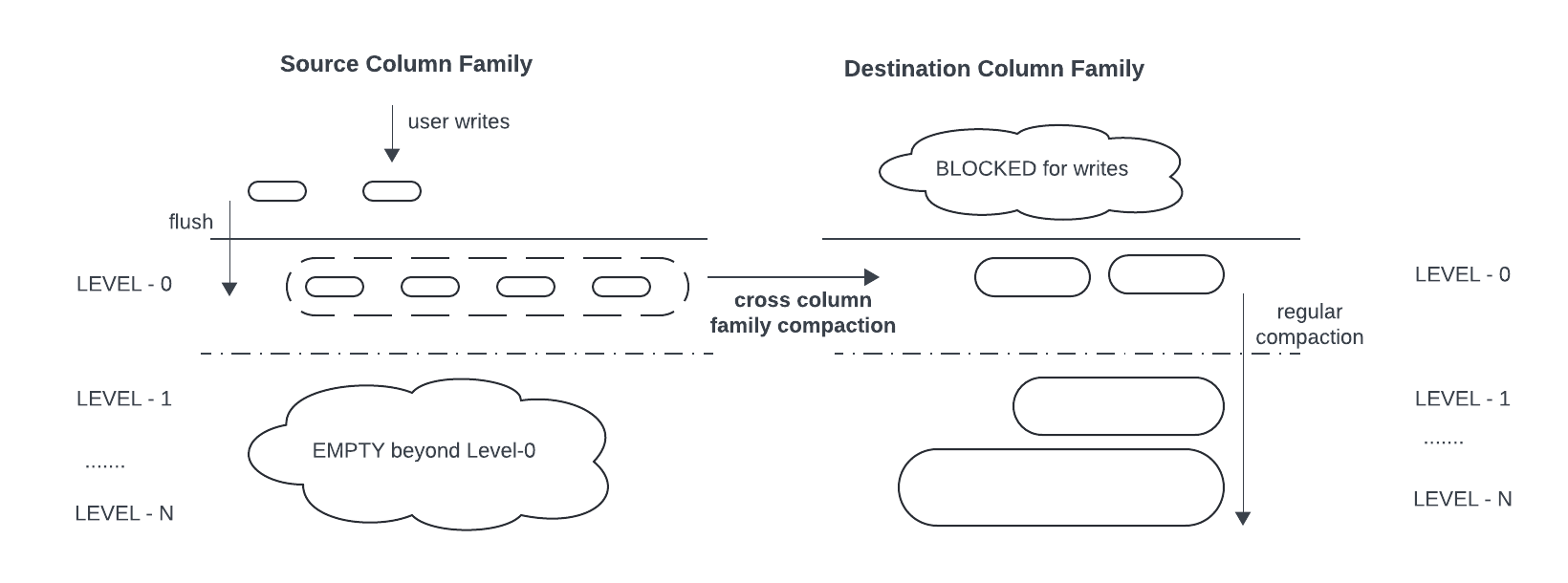}
    \caption{In a cross-column-family compaction, the user-facing column family serves as the source, receiving data from the application, while the internal destination column family receives this data through compaction and continues the compaction.}
    \label{fig:cross_column_family_compactin}
\end{figure}

Figure \ref{fig:cross_column_family_compactin} illustrates an example of cross-column-family compaction. In this example, the source column family is user-facing and accepts user writes into its in-memory skip-list. Once the skip-list is full, it is flushed to the source column family's Level-0 as an SST file. When the number of SST files in Level-0 reaches a configurable threshold, a compaction job is dispatched. Instead of compacting data into lower levels within the same column family, the job writes the compacted output to the Level-0 of the destination column family. The destination column family is protected from external writes as it is only used for backend compaction purposes and is not user-facing. In this design, all levels beyond Level-0 in the source column family remain empty.

The destination column family can either function like a regular column family, compacting the data it receives at Level-0 into lower levels, or it can act as a source column family for yet another destination column family. In the latter case, the destination column family only compacts data from its Level-0 into its own destination column family, leaving levels beyond Level-0 empty.

\subsection{Tierveling}
Compaction is the process of merging data from smaller SST files into larger ones. There are two primary merging strategies: tiering and leveling. The trade-offs between these approaches are well studied, with a key difference being that in tiering, each level contains multiple sorted runs, whereas in leveling, all levels except Level-0 maintain a single sorted run. In transformation-embedded LSM-trees, we adopt a hybrid merging approach called tierveling. Under this approach, leveling is used when the identity function (pure compaction) is applied, while tiering is employed for all other transformation functions. This distinction arises because the identity function preserves the table structure, enabling each level of the LSM-tree to maintain a single sorted run, which optimizes read performance. However, when transformations are applied, whether immediately or gradually, the data transitions from one table to another, making leveling unsuitable for these scenarios.


\begin{figure}[H]
    \centering
    \includegraphics[width=\linewidth]{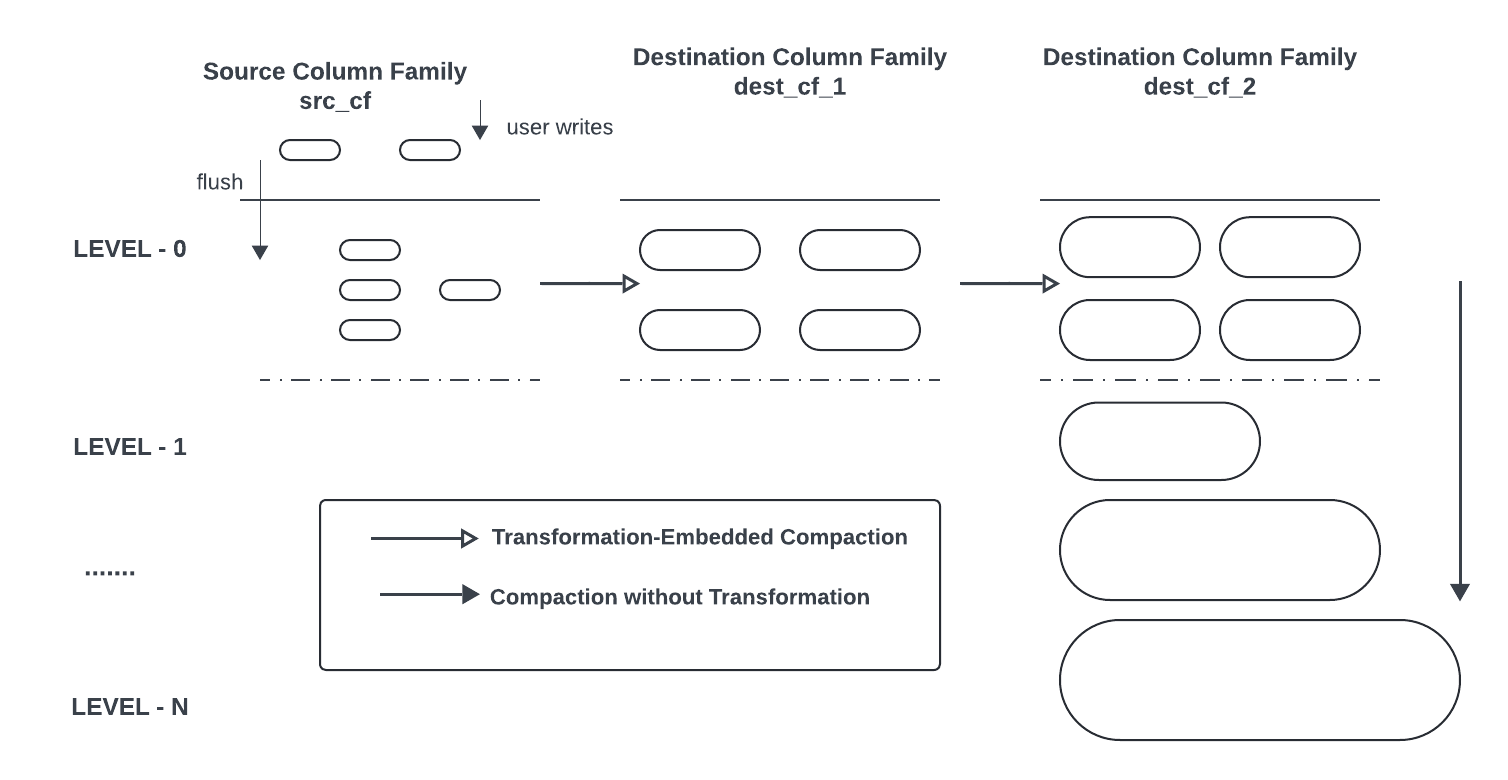}
    \caption{In Tierveling, LSM-trees with embedded transformations perform the compaction with Tiering strategy, while the LSM-trees that have no transformations embedded compact with Leveling strategy.}
    \label{fig:tierveling}
\end{figure}

Figure \ref{fig:tierveling} illustrates the mixed tiering and leveling (tierveling) flow within the logical LSM-tree for the source column family src\_cf. In this example, the compaction jobs for both the source column family (src\_cf) and the destination column family (dest\_cf) are transformation-embedded. Their outputs are written into another column family, with data in Level-0 only. Consequently, the sorted runs in these column families overlap and follow the tiering approach. In contrast, the last column family (dest\_cf\_2) is not transformation-embedded and functions as a traditional LSM-tree. It engages in regular compaction, maintaining a single sorted run in all levels beyond Level-0. Thus, Figure \ref{fig:tierveling} represents the logical LSM-tree structure for the column family \texttt{src\_cf}.

\subsection{Transformation Algebra}
Transformers are composable, and they possess the following attributes:
\begin{itemize}[itemsep=0pt, topsep=0pt, left=0pt]
    \item \textbf{Associativity}: grouping transformers applied to a column family in different ways produces the same final result.
    \begin{equation}
        (F(Tr_a) + F(Tr_b)) + F(Tr_c) = F(Tr_a) + (F(Tr_b) + F(Tr_c))
    \end{equation}
    \item \textbf{Commutativity}: the order in which the transformation functions are applied does not affect the final result.
    \begin{equation}
        F(Tr_a) + F(Tr_b) = F(Tr_b) + F(Tr_a)
    \end{equation}
\end{itemize}

\subsection{Cost Model}
We develop a cost model for transformation-embedded compaction (TEC) to predict whether a given transformation is beneficial by evaluating its potential write overhead and read performance improvement. Our analysis examines the cost from four key perspectives: write throughput, point queries, range queries, and space amplification. Due to space constraints, the detailed cost model is provided in Appendix \ref{appendix:cost-model} for interested readers.   

\section{Implementation of MyCelium}
 We now describe the implementation of Mycelium, our storage engine built on top of RocksDB. Mycelium extends RocksDB to streamline data transformations into the background compaction process.

 \subsection{Mycelium Architecture}
Figure \ref{fig:mycelium-architecture} illustrates the high-level architecture of Mycelium, along with an example code path for opening the database and writing to a column family. The core object structs for Mycelium are the MemTable, the SST files, and the Write Ahead Log (WAL). The components that have been modified to support transformation-embedded compaction are highlighted in green, which includes: 1, the \textbf{Options} to manage the configuration parameters; 2, the \textbf{ColumnFamilyData} for creating source and destination families for storing the transformed outputs; 3, the \textbf{Transformer} for implementing the transformation functions; 4, the \textbf{Compaction} for invoking transformations while compacting and for installing the compaction result into the destination column families. 

\begin{figure}[H]
    \centering
    \includegraphics[width=\linewidth]{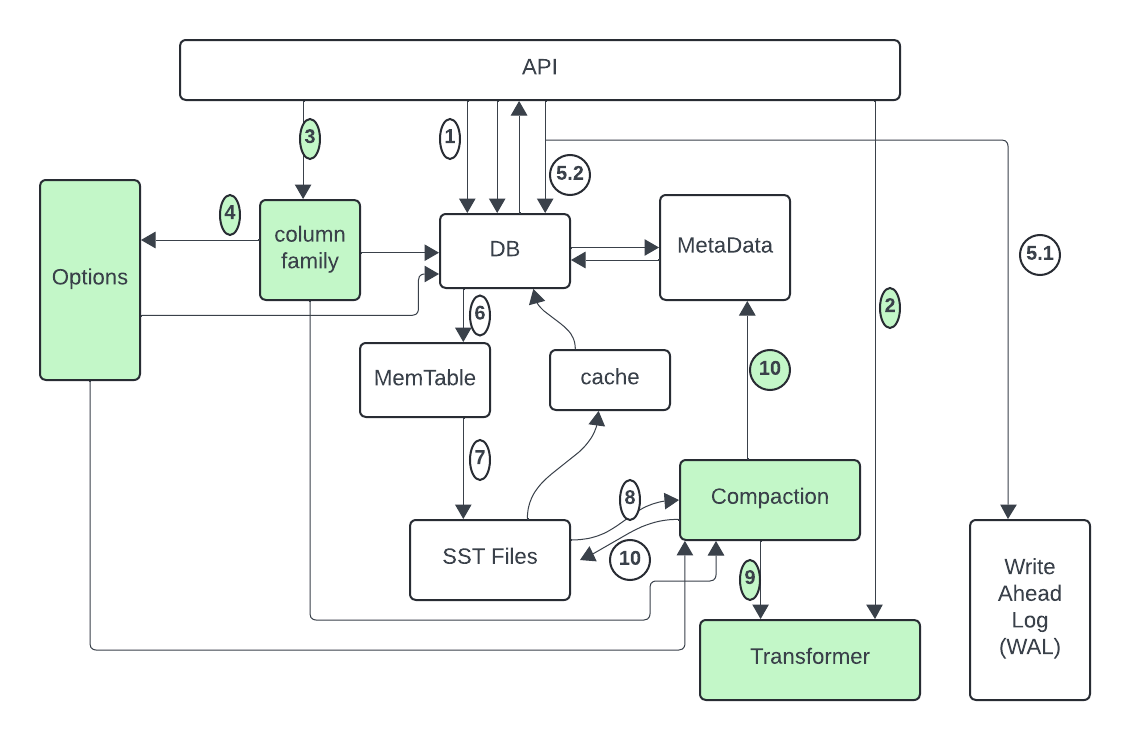}
    \caption{High-level architecture of Mycelium}
    \label{fig:mycelium-architecture}
\end{figure}

Below, we outline the steps in the code path as marked in the figure:
\ding{172} Opening the database
\ding{173} Creating transformers
\ding{174} Creating column families to store transformed outputs
\ding{175} Storing configuration options for column families
\ding{176} 5.1 Writing to the Write-Ahead Log (WAL); 5.2 Writing to a column family
\ding{177} Buffering writes into the MemTable
\ding{178} Flushing MemTables to disk as SST files
\ding{179} Compaction job preparing the input
\ding{180} Invoking Transformer to produce transformed outputs
\ding{181} Installing compaction result by writing new SST files and updating metadata

\subsection{Transformers}
Mycelium introduces a \texttt{Transformer} interface, allowing users to supply custom transformation functions. In addition, it provides three built-in implementations, each performing a distinct type of transformation. We outline the interface and the implementations in more details below:

\subsubsection{Function Interface} A Transformer function interface unifies different types of transformations, which the three built-in implementations implement. The following methods are to be implemented for adding a custom transformer:

\textbf{Prepare()}: Prepares the transformer for the next compaction job by granting \texttt{lock} to the compaction job, and clearing the staging area. As transformers are associated with a column family, at any point of time, only one compaction job can have access to the transformer.

\textbf{Transform(input, outputs)}: Converts the input of a key-value pair (k, v) into a vector of key-value pair outputs (k’, v’). The transformer takes the output of the identity function (compaction) as the input, performs the transformation, and returns the transformed outputs.

\textbf{Retrieve()}: Retrieves the staged transformation outputs. Staged transformation outputs are in the format of (key, [value]). After retrieval the lock is released.

\subsubsection{Split} Splitting a column group into smaller ones transforms row-wise data into columnar data, optimizing it for future analytical queries. This process can occur either gradually, over multiple rounds of compactions, or immediately, in a single round, until the data reaches its final destinations in the designated column families. Through creating destination column families, one can customize how a split-transformer splits column groups of source column families. By definition, splitting typically results in more than two destination column families, often a power of two, but the exact number is determined by the customization of the transformer. Figure \ref{fig:split-transformer} provides an example of how a split-transformer gradually divides the column group of a source column family (\texttt{src\_cf}) by compacting data into successive column families.

\begin{figure}
    \centering
    \includegraphics[width=\linewidth]{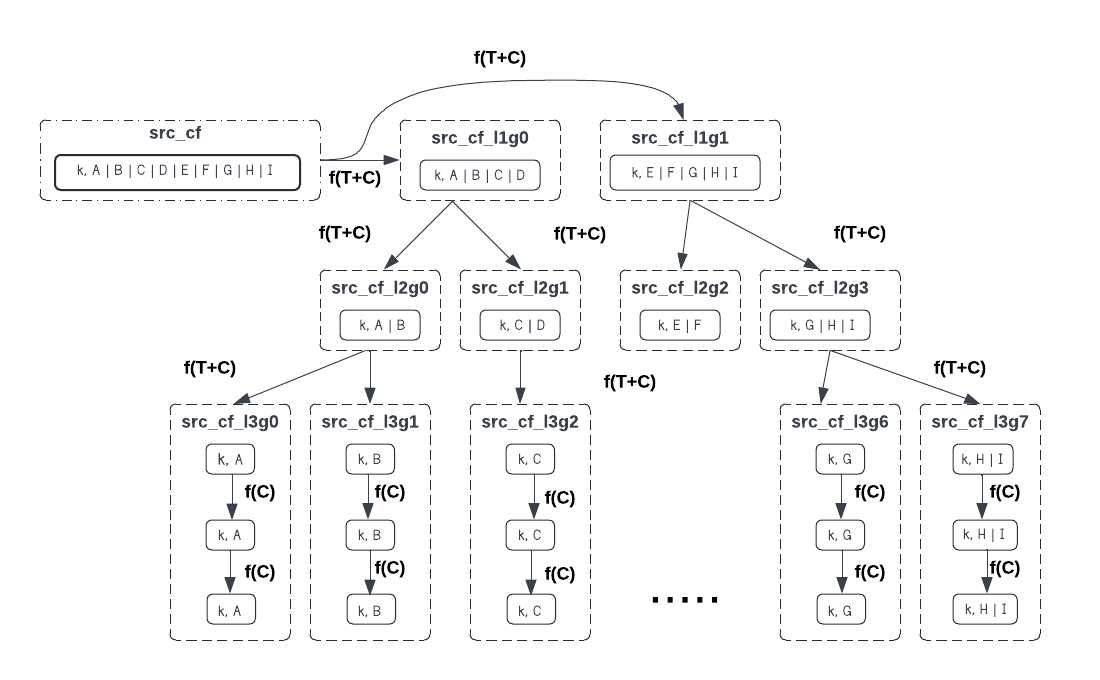}
    \caption{Split transformer splits column groups while compacting}
    \label{fig:split-transformer}
\end{figure}
\vspace{-2mm}

In this example, data initially enters the storage system into the source column family (\texttt{src\_cf}), which contains nine columns (A|B|C|D|E|F|G|H|I). Over the course of their life cycle, the data undergoes five compacting movements until reaching their final destinations. The first compaction splits the column group into two smaller groups: one with four columns (A|B|C|D), compacted into the column family \texttt{src\_cf\_l1g0}, and the other with five columns (E|F|G|H|I), compacted into the column family \texttt{src\_cf\_l1g1}.

This progressive splitting and compaction process continues until the original column group is divided into eight smaller column groups. Of these, seven groups each contain a single column, while one group contains two columns. The first three data movements involve both splitting and compaction, denoted as \texttt{f(C+T)}, where \texttt{C} represents compaction and \texttt{T} represents transformation. In contrast, the final two movements involve only compaction without any transformation, denoted as \texttt{f(C)}. These last two compactions occur inside the final destination column families.

\subsubsection{Convert} A convert-transformer changes the file format in which the \texttt{value} in key-value pairs is stored. Examples of such transformations include converting data from CSV to JSON, text to compressed formats, JSON to Protobuf, Protobuf to FlatBuffers, and other similar format conversions. Unlike split-transformers, which progressively alter data over multiple stages, convert-transformers apply the format transformation immediately during compaction, transferring data from the source column family to a system-created column family designated for the converted data. Typically, the number of destination column families in the case of convert-transformer is one.

Figure \ref{fig:convert-transformer} illustrates an example of this process. In the example, data in the source column family (\texttt{src\_cf}) is transformed and stored in the destination column family (\texttt{src\_cf\_converted}). Here, \texttt{src\_cf} has a convert-transformer defined, enabling the transformation-embedded compaction (f(T+C)) that applies both transformation (T) and compaction (C) in a single job. In contrast, \texttt{src\_cf\_converted} does not have a transformer defined. Once the data reaches \texttt{src\_cf\_converted}, all subsequent movements occur entirely within this column family, with no further transformations applied.

\begin{figure}
    \centering
    \includegraphics[width=\linewidth]{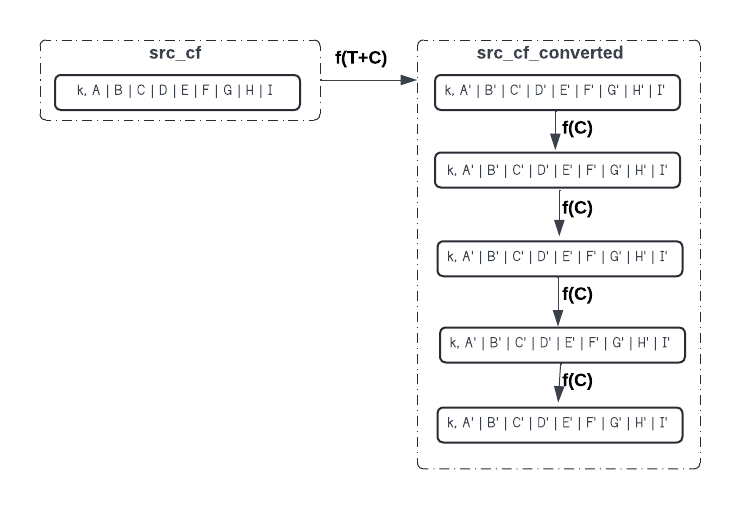}
    \caption{Convert transformer converts value format while compacting}
    \label{fig:convert-transformer}
\end{figure}

\subsubsection{Augment} An augment-transformer enhances the source data by creating additional data structures, such as secondary indexes, materialized views, or other auxiliary structures. Typically, the number of destination column families for augment-transformers is at least two, with no theoretical limit on how many auxiliary structures can be generated through compaction. Figure \ref{fig:augment-transformer} illustrates an example of creating a secondary index.

In this example, data is initially flushed into the column family \texttt{src\_cf}. When compaction begins, the compacted outputs are written into \texttt{src\_cf\_primary}, while a secondary index on column A is created in the column family \texttt{src\_cf\_secondary\_1}. Although no format or structural changes are applied to the source data, the compacted outputs are redirected from the source column family (\texttt{src\_cf}) to an internally-created destination column family (\texttt{src\_cf\_primary}). By adopting this approach, the source column family (\texttt{src\_cf}) retains data only in its Level 0, with all higher levels remaining practically empty. This design ensures that the primary data in \texttt{src\_cf\_primary} and the secondary index in \texttt{src\_cf\_secondary\_1} maintain consistent data placement across their respective levels. 

Similar to convert-transformers, augment-transformers only apply transformations to the original source column families that receive the flushed MemTables, as these are the column families where the transformers are defined. Internally-created column families, on the other hand, perform regular compactions without applying any additional transformations to the data.

\begin{figure}[h]
    \centering
    \includegraphics[width=\linewidth]{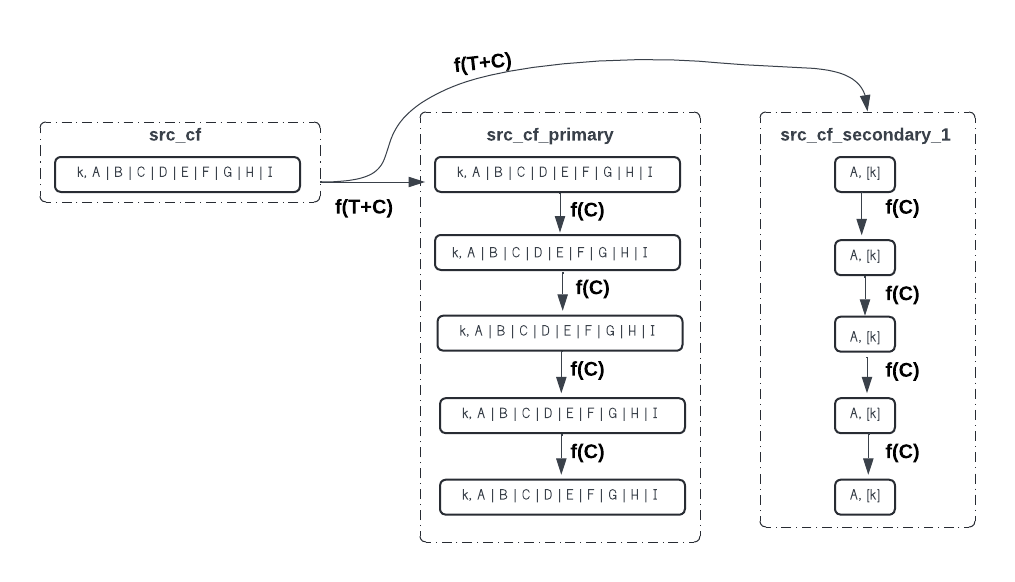}
    \caption{Augment transformer creates secondary index while compacting}
    \label{fig:augment-transformer}
\end{figure}

\subsubsection{Transformer Algebra}
Transformers can be overlaid on a column family, enabling multiple types of transformations to be defined and executed during compaction. However, overlaying transformations does not mean that multiple transformations are applied simultaneously in a single compaction attempt. To balance the workload and prevent overburdening any single compaction, Mycelium enforces the following rules for assigning transformers to column families:

\begin{enumerate}[itemsep=0pt, topsep=0pt]
    \item At most one transformer per column family: Each individual column family, whether user-facing or internally created, can have at most one transformer associated with it.
    \item At most one gradual transformer per logical column family: A logical column family is defined as the sum of the user-facing column family and all the internally created column families used to store transformed outputs for it. While multiple transformations can be defined across the logical column family, only one gradual transformer can be associated with it.
    \item Gradual transformers are applied first: When a column family has both gradual and non-gradual transformations associated with its logical column family, the gradual transformer is applied before other types of transformations.
\end{enumerate}

Table \ref{tab:transformer-policy} illustrates this policy with an example of a logical column family supported by a six-level logical LSM-tree. A logical LSM-tree refers to the conceptual representation of the LSM-tree structure, encompassing all data transformations applied across its levels. It includes both the user-facing column family and the internally created column families, each of which is backed by its own physical LSM-tree. In this example, two types of data transformations are applied. First, a gradual transformation splits data from a row-oriented format to a column-oriented format, applied across logical levels 0 to 2 (referred to as the "Global Level" in the table). Second, a non-gradual transformation converts the data format from Protobuf to FlatBuffers, applied across logical levels 2 to 3. No transformations are performed in logical levels 3 to 5.  In the table, the levels of the physical LSM-tree are referred to as the "Local Level". 

\begin{table}[h!]
\centering
\resizebox{\columnwidth}{!}{
\begin{tabular}{|@{}c|l|c|c|c@{}|}
\toprule
\parbox[t]{2cm}{\centering \textbf{Logical} \\ \textbf{LSM-tree} \\ \textbf{Level}} & \parbox[t]{2cm}{\centering \textbf{Column} \\ \textbf{Family} \\ \textbf{Name}} & \parbox[t]{2cm}{\centering \textbf{Column} \\ \textbf{Family} \\ \textbf{Type}} & \parbox[t]{2cm}{\centering \textbf{Local} \\ \textbf{Level}} & \parbox[t]{2cm}{\centering \textbf{Transformer} \\ \textbf{Type}} \\
\midrule
0 & my\_cf\_json & user-facing & 0 & splitting \\
1 & my\_cf\_l1g[0..1]\_json & internal & 0 & splitting \\
2 & my\_cf\_l2g[0..3]\_json & internal & 0 & converting \\
3 & my\_cf\_l2g[0..7]\_FlatBuffers & internal & 0 & none \\
4 & my\_cf\_l2g[0..7]\_FlatBuffers & internal & 1 & none \\
5 & my\_cf\_l2g[0..7]\_FlatBuffers & internal & 2 & none \\
\bottomrule
\end{tabular}
}
\caption{Example showing Mycelium's transformer policy}
\label{tab:transformer-policy}
\end{table}
\vspace{-1mm} 

\subsection{Writes}
Since the transformations embedded in Mycelium operate exclusively in the background during compaction, writes in Mycelium function the same way as in standard RocksDB.

\subsection{Reads}
Transformations in Mycelium are designed to optimize future read queries, allowing most queries to naturally and directly benefit from the transformed data without requiring explicit actions. Examples of queries that benefit the most from Mycelium's transformations include:
\begin{enumerate}[itemsep=0pt, topsep=0pt]
    \item Analytic queries that access only one or a very small number of columns.
    \item Queries performing numeric operations.
    \item Index-based queries.
\end{enumerate}
However, certain types of transformations, such as column group splitting, present additional challenges. When row-wise data is split and stored in a columnar format, reconstructing the original rows requires merging the split columns. To address this, Mycelium implements a column merge operator, enabling seamless row reconstruction for these scenarios.

\subsection{Algorithms}
We include in Appendix ~\ref{appendix:algo} the core algorithms used in Mycelium, for anyone who is interested, which includes three key components: 1. An API exposed to users, which sets up the transformers and creates the internal destination column families, and upon invocation initiates the transformation-embedded compaction. 2. The function responsible for applying a transformation to a column family that has an associated transformer. 3. Installing compaction results into the manifest files, which enables the transfer of data from the source column family to the designated destination column families.

\section{Evaluation}
We conduct various experiments to measure the performance of Mycelium. Specifically we evaluate the write throughput, and the write and read latencies. We hope that the tests can provide answers to address the following key questions: 1. Are Mycelium transformations effective cost amortization? 2. How beneficial are these transformations for future queries?

\subsection{Hardware Setup}
We deploy Mycelium on a Linux machine running 64-bit Ubuntu 20.04.6 LTS, equipped with 2 CPUs each having 16 cores clocked at 3.00 GHz. It has 128 GB of ECC memory and two 480 GB 6G SATA SSDs.

\subsection{Databases Tested}
We evaluate and compare the write throughput and the read and write performance of four flavors of Mycelium databases, each implementing a different type of data transformation, against RocksDB-based baselines. 

\subsubsection{Baselines} Below, we outline the configurations used as baselines. The first baseline represents standard RocksDB, which does not perform any data transformations during the compaction process. It is used to evaluate both write throughput and read latencies. The second through fourth baseline each apply a different type of data transformation outside the compaction process and are used exclusively for comparing the write throughput with the Mycelium configurations that perform transformations within the compaction process.

\begin{itemize}[itemsep=0pt, topsep=0pt, left=0pt]
    \item \textbf{Baseline}: A standard RocksDB key-value store.
    \item \textbf{Baseline-Splitting}: Incoming row-wise data with 32 columns is divided into smaller groups of 4 columns each, and each group is written to a separate column family.
    \item \textbf{Baseline-Converting}: The data format is converted from JSON to FlatBuffers before being written into the column family.
    \item \textbf{Baseline-Augmenting}: A secondary index is actively maintained alongside the primary data as it is written into the column family.
\end{itemize}

\subsubsection{TE-LSMs} We evaluate five TE-LSMs: three implement a single type of transformer available in Mycelium, while the fourth demonstrates the algebraic composition of two transformers, showcasing the capabilities of transformer algebra. Additionally, we include a TE-LSM that applies the identity function, i.e. a transformation no-op.

\begin{itemize}[itemsep=0pt, topsep=0pt, left=0pt]
    \item \textbf{Mycelium-Splitting}: A \textsc{split} transformer is embedded to partition row-oriented data of 32 columns gradually into smaller column-oriented groups of 4 columns during compaction, each time evenly splitting into two column groups.
    \item \textbf{Mycelium-Converting}: A \textsc{convert} transformer is embedded to transform data arriving in JSON format into FlatBuffers format during compaction.
    \item \textbf{Mycelium-Augmenting}: An \textsc{augment} transformer is embedded to maintain a secondary index during compaction. With this configuration, data in all levels beyond Level-0 is indexed. Given an LSM-tree size factor of 10, over 99.99\% of the data is indexed. 
    \item \textbf{Mycelium-Split-Converting}: Two transformers, \textsc{split} and \textsc{convert}, are embedded. The \textsc{split} transformer divides the incoming data in the same way as the transformer in \texttt{Mycelium-Splitting} configuration. The \textsc{convert} transformer transforms the data format from Protobuf to FlatBuffers at the end of the splitting process.
    \item \textbf{Mycelium-Identity}: An \textsc{identity} transformer is embedded, applying the identity function during compaction as a no-op transformation.
\end{itemize}

\subsection{Workloads} 
We run our experiments using three workloads—\textbf{write}, \textbf{read}, and \textbf{scan}—within the YCSB framework. The write workload evaluates write throughput, while the read and scan workloads measure read performance. All workloads are executed on pre-loaded test beds containing 100GB of data. This ensures that, for both Mycelium and Baseline configurations, every level of the LSM-tree is populated with data, resembling the LSM-trees in operation.

\subsubsection{Queries} Using SQL notation, we describe the queries that are executed in the workloads as follows. In this notation, T, K, and V stand for the column family (table), the column name for the keys, and the column name for the row-wise values. $V_1, V_2, ..., V_i$ stand for the column names for the columns inside the rows, k, v, $v_1, v_2, ..., v_i$ are the values for keys, row values, and column values.

\begin{enumerate}[itemsep=0pt, topsep=0pt]
    \item \textbf{Q1(INSERT)}:
    \texttt{INSERT INTO T VALUES (k, v);}
    
    \item \textbf{Q2(Range-Query-Read-Column)}:
    \texttt{SELECT MAX($V_i$) FROM T WHERE K $\geq$ k1 \allowbreak AND K $<$ k2;}
    
    \item \textbf{Q3(Point-Query-Read-Column)}: 
    \texttt{SELECT $V_i$ FROM T WHERE K = k;}
    
    \item \textbf{Q4(Non-Key-Predicate-Range-Query-Read-Column)}: 
    \texttt{SELECT MAX($V_i$) FROM T WHERE $V_i$ $\geq$ $v_{i1}$ \allowbreak AND $V_i$ $<$ $v_{i2}$;}
    
    \item \textbf{Q5(Non-Key-Predicate-Point-Query-Read-Row)}: 
    \texttt{SELECT * FROM T WHERE $V_i$ = $v_i$;}
    
    \item \textbf{Q6(Range-Query-Read-Row)}:
    \texttt{SELECT * FROM T WHERE K $\geq$ k1 AND K $<$ k2;}
    
    \item \textbf{Q7(Point-Query-Read-Row)}:
    \texttt{SELECT * FROM T WHERE K = k;}
\end{enumerate}

\subsubsection{Test Data} We generate synthetic test data to evaluate the workloads under the eight configurations described earlier. The details of the synthesized data are as follows:
\begin{itemize}[itemsep=0pt, topsep=0pt, left=0pt]
    \item Keys are generated uniformly as numeric values and then converted into 16-byte-long strings.
    \item Each value consists of 50 columns, with each column containing either a 24-byte string or a randomly generated unsigned 64-bit integer.
    \item Two data formats are used: Protobuf for configurations without conversion transformations, and JSON for those involving conversion transformations. Baseline-1 is tested with both formats, while Baseline-2 and Baseline-4 use Protobuf, and Baseline-3 uses JSON.
\end{itemize}

\subsection{Mycelium Performance}
We present the results of our experiments evaluating Mycelium's performance, organized to address the questions introduced at the beginning of this section, which will serve to confirm or refute our hypothesis.

\subsubsection{\textbf{Question 1: Are Mycelium transformations effective cost amortization?}\\}

Performing additional data processing inevitably introduces system overhead, potentially affecting performance. In Mycelium style databases, this overhead arises from integrating data transformations with background compactions. We measure this overhead as a reduction in write throughput compared to a baseline system, using the reduction percentage as a metric to evaluate Mycelium's effectiveness in absorbing the cost of transformation work. Our experimental results show that Mycelium incurs a write throughput reduction ranging from 10.09\% to 21.25\% (Table \ref{tab:xput} ) for performing the data transformations.

\setlength{\belowcaptionskip}{0pt} 
\begin{table}[H]
    \centering
    \resizebox{\columnwidth}{!}{  
        \begin{tabular}{|l|r|}
            \hline
            \textbf{TE-LSM Configurations}      &  \textbf{Throughput Penalty} \\ \hline
            TE-LSM-Splitting  & 10.93\% \\
            Baseline-Splitting    & 57.98\%  \\ \hline
            TE-LSM-Converting & 15.42\% \\
            Baseline-Converting & 38.03\% \\ \hline
            TE-LSM-Augmenting & 10.09\% \\
            Baseline-Augmenting & 34.84\% \\ \hline
            TE-LSM-Split-Converting & 21.25\% \\ \hline
            TE-LSM-Identity  & -5.25\% \\ \hline
        \end{tabular}
    }
    \caption{Write Throughput Overhead Relative to RocksDB Baseline for Mycelium and Naive Approaches}
    \label{tab:xput}
\end{table}
\vspace{-1mm}

Before starting the write throughput experiments, we pre-load into each LSM-tree 100 GB of data, consisting of 200,000,000 records, each approximately 500 bytes in size. Given the storage capacity configured for our test databases, this ensures that every level of the LSM-tree is populated with data. To begin the experiments, we first tune the configuration parameters for the baseline RocksDB on our chosen hardware setup to maximize throughput. The same set of parameters (Appendix~\ref{appendix:config}) is then applied to all other databases. Finally, we run Q1 with 8 concurrent clients for 20 minutes per test. 

The experimental results validate the first half of our hypothesis: the Mycelium approach effectively hides the transformation work while incurring only a small overhead during compaction. Specifically, the write throughput overhead for Mycelium LSM-trees performing a single transformation remains below 16\%, and around 20\% when performing two transformations simultaneously. These results also highlight the significant efficiency advantage of the Mycelium approach compared to the naive approach, where transformations result in a 35\% to 60\% reduction in throughput.

Interestingly, the LSM-tree with Mycelium-Identity, which applies a no-op transformation, achieves a 5\% higher throughput than the RocksDB baseline. Although counterintuitive, this improvement is attributed to Mycelium’s hybrid merging policy, Tierveling, which employs a Tiering strategy to efficiently move data out of Level-0 in the user-facing LSM-tree. This strategy helps mitigate write stalls, thereby enhancing write efficiency. However, this improvement comes with a trade-off: it may lead to higher read costs. In a stable LSM-tree, where the number of deletions equals the number of insertions, this advantage should diminish, and the write throughput of Mycelium-Identity is expected to align closely with the baseline.

\subsubsection{\textbf{Question 2: How beneficial are these transformations for future read queries?}}
\begin{figure}[ht]
    \centering

    \begin{subfigure}[b]{0.48\columnwidth}
        \centering
        \includegraphics[width=\columnwidth]{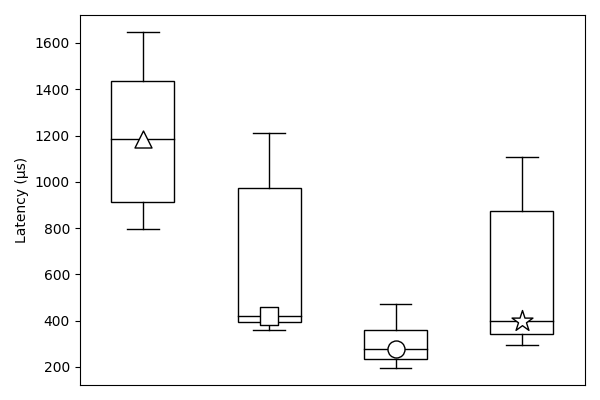}
        \caption{Range Query (Q2)}
        \label{fig:Q2}
    \end{subfigure}
    \hfill
    \begin{subfigure}[b]{0.48\columnwidth}
        \centering
        \includegraphics[width=\columnwidth]{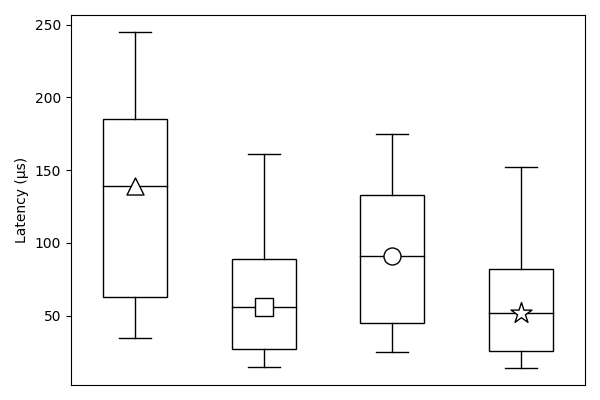}
        \caption{Point Query (Q3)}
        \label{fig:Q3}
    \end{subfigure}

    \vspace{-1mm}  
    \includegraphics[width=0.5\textwidth]{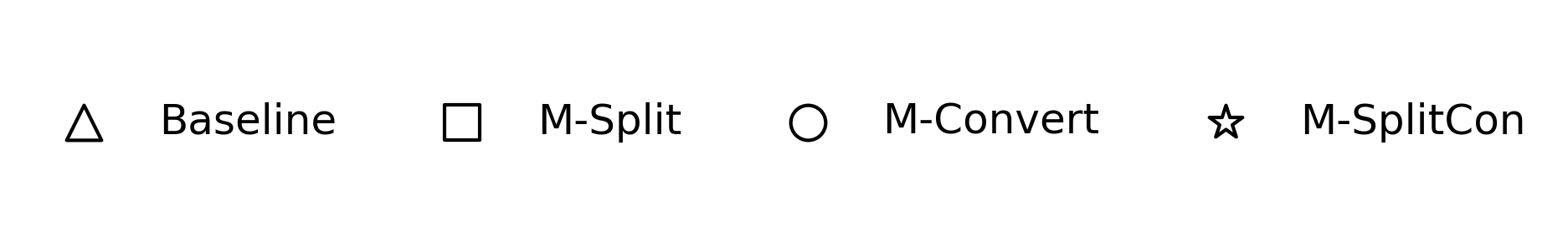}
    \vspace{-1cm}
    \caption{Column-retrieving queries benefit significantly from both split and convert transformations, as the split transformation reduces I/O and deserialization costs by minimizing record size, while the convert transformation compresses data into much more space-efficient format (FlatBuffers). Q2 achieves better performance in TE-LSM than Q3, due to the amplified effect of the benefit in a range query as opposed to the point query.}
    \label{fig:col-read-queries}
\end{figure}

The answers to this question validate the second half of our hypothesis: embedding data transformations within compaction makes LSM-trees more read-optimized. We go through each transformation type to evaluate the performance of Q2 through Q7, as detailed in the following sections. For each query, the keys used in the predicates are generated using a "zipfian" distribution. Each query is executed 100,000 times in a batch, and the batch is repeated multiple times to warm up the cache. During latency collection, metrics such as min, P25, P50, P75, and max are recorded.

\vspace{4mm}
\begin{figure}[ht]
    \centering

    \begin{subfigure}[b]{0.48\columnwidth}
        \centering
        \includegraphics[width=\columnwidth]{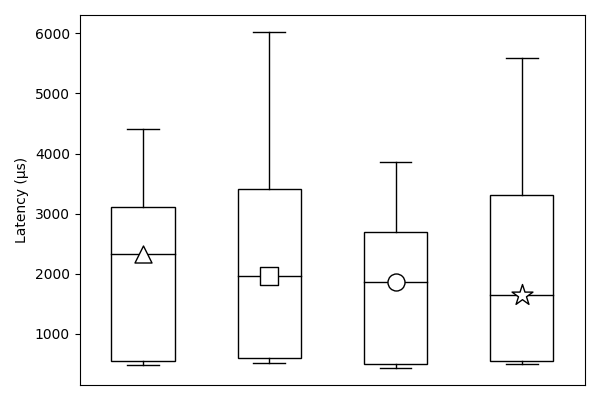}
        \caption{Range Query (Q6)}
        \label{fig:Q6}
    \end{subfigure}
    \hfill
    \begin{subfigure}[b]{0.48\columnwidth}
        \centering
        \includegraphics[width=\columnwidth]{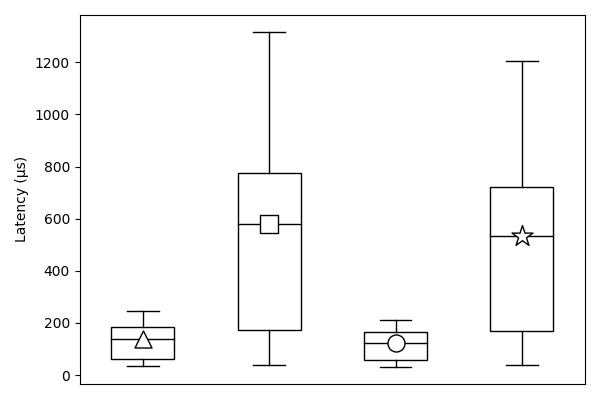}
        \caption{Point Query (Q7)}
        \label{fig:Q7}
    \end{subfigure}

    \vspace{-1mm}  
    \includegraphics[width=0.5\textwidth]{legend_horizontal.png}
    \vspace{-1cm}
    \caption{Row-retrieving queries showing negative impact from the split transformation due to the need to reassemble the split records to reconstruct the row. While the convert transformation achieves slightly lower latencies than the baseline, the split transformation has higher latency across most metrics in Q7, and mixed performance with higher P100 but lower P50 in Q6.}
    \label{fig:row-read-queries}
\end{figure}

\textbf{Split-Transformation} Split transformations improve the performance of queries that access a small subset of columns but degrade the performance of row-based queries. As expected, compared to the RocksDB baseline, the TE-LSM with split transformations performs better on column-retrieving queries such as Q2 (Figure~\ref{fig:col-read-queries}(a)) and Q3 (Figure~\ref{fig:col-read-queries}(b)), but worse on row-retrieving queries like Q6 (Figure~\ref{fig:row-read-queries}(a)) and Q7 (Figure~\ref{fig:row-read-queries}(b)).

Q2 and Q3 retrieve only single columns, allowing them to fully leverage the significantly reduced split record sizes. This leads to lower I/O and deserialization costs. Notably, Q2 (a range query) benefits even more than Q3 (a point query) due to the amplified effect of smaller I/O costs over a range of data. Q2 achieves a 2.8x read latency improvement compared to the baseline. Conversely, row-based queries like Q6 and Q7 suffer because the previously split records must be reassembled, introducing additional overhead. Q7 performs worse than Q6 since Q6, being a range query, gains more from the smaller record sizes, partially offsetting the cost of reassembling columns.

\textbf{Convert-Transformation} The TE-LSM with conversion transformations consistently outperforms the baseline across all four queries—Q2, Q3 (Figure~\ref{fig:col-read-queries}, Q6, and Q7 (Figure~\ref{fig:row-read-queries}). In our evaluation, this transformation converts data from the input JSON format to the more space-efficient FlatBuffers format, resulting in a 34.76\% reduction in SST file size. The smaller file size directly translates to lower I/O costs. Among the queries, Q2 shows the most significant performance gain because it calculates the maximum value over a range of values, making it particularly well-suited to benefit from the compact numeric representation of the values in the converted format. As a result, all queries experience reduced latency, with Q2 achieving the highest improvement—a 4.25× reduction in query latency.

\vspace{-1mm}

\begin{table}[ht]
    \centering
    \resizebox{\columnwidth}{!}{  
        \begin{tabular}{|l|c|c|c|c|}
            \hline
            \multirow{2}{*}{Configuration} & \multicolumn{2}{c|}{Point Query} & \multicolumn{2}{c|}{Range Query} \\
            \cline{2-5}
                                       & P50 (µs) & P99 (µs) & P50 (µs) & P99 (µs) \\
            \hline
            TE-LSM-Augmenting              & 243  & 751 & 3049 & 7803 \\
            \hline
            Baseline-Augmenting                       & >3 min & >5 min & >30 min & >1 hr \\
            \hline
        \end{tabular}
    }
    \caption{Index Queries Performance Comparison}
    \label{tab:index-query}
\end{table}
\vspace{-1mm}

\textbf{Augment-Transformation} The augment transformation used in our evaluation—specifically for creating a secondary index—significantly improves the performance of index queries Q4 and Q5 (Table \ref{tab:index-query}). Since RocksDB lacks native support for secondary indexes, queries on non-key columns must perform full table scans. As a result, Q4 and Q5 suffer from severe delays, with the P50 latency of the point query extending to several minutes and the P50 latency of the range query exceeding 30 minutes. Applying the augment transformation yields performance improvements for Q4 and Q5 that are at least several hundred thousand times greater than the baseline.

\textbf{Split-Convert-Transformation} The TE-LSM used in our evaluation applies algebraic operations that combine both split and convert transformations. Queries Q2, Q3, Q6, and Q7 perform similarly—but slightly better—under this combined transformation compared to the split transformation alone. This slight improvement is primarily driven by the dominant impact of the split transformation, with additional benefits from the convert transformation. Compared to the baseline, Q2 and Q3 (Figure \ref{fig:col-read-queries}) show noticeable performance improvements. In contrast, Q6 (Figure \ref{fig:row-read-queries}(a) exhibits mixed results, with a lower P50 latency but a higher P99 latency, while Q7 (Figure \ref{fig:row-read-queries}(b)) experiences worse overall performance. 

\section{Related Work}

\textbf{Compaction optimizations}: There has been extensive research focused on optimizing compaction since the inception of LSM-trees. These efforts encompass a variety of strategies aimed at improving efficiency and reducing overhead. For example, dCompaction \cite{pan2017dcompaction} reduces compaction frequency by postponing and combining some compactions with subsequent ones. TRIAD\cite{balmau2017triad} minimizes disk writes by retaining frequently updated keys in memory, while VT-tree \cite{shetty2013building} employs a stitching mechanism to avoid unnecessary copying of input pages. The "spring-and-gear" merge scheduler in bLSM\cite{sears2012blsm} mitigates write stalls by bounding write latency. Pipelined merge \cite{zhang2014pipelined} exploits I/O parallelism and hardware capabilities, WiscKey \cite{lu2017wisckey} decouples the storage of keys from values to minimize data movement, and PebblesDB \cite{raju2017pebblesdb} partitions the key space into separate trees for more efficient data organization. 

While these compaction optimizations primarily aim to reduce overhead and improve efficiency--either through internal process refinement or better hardware utilization--our research on TE-LSM and Mycelium, in contrast, adopts a fundamentally different approach. Rather than viewing compaction solely as a "necessary evil" to be mitigated, we treat it as a strategic opportunity to embed additional work whose cost can be amortized thereby enhancing the overall system performance. This approach is viable for IO-bound data transformations that can share the IO required by compaction.

\textbf{Data transformations}: Recent research on data transformation and data organization in storage systems has increasingly focused on optimizing data for mixed workloads \cite{arulraj2016bridging}\cite{psaroudakis2015scaling}\cite{lyu2021greenplum}. For example, Relational Memory \cite{roozkhosh2021relational} transparently transforms data from a row-wise format into arbitrary column groups during query execution, improving flexibility and performance. ${H_2O}$ \cite{alagiannis2014h2o} employs multiple storage layouts and dynamically generates execution plans based on access paths to select the most suitable format, continuously adapting to evolving access patterns. Doraemon \cite{tang2019learned} leverages machine learning to build adaptive indexes by caching trained models and enriching training data with access frequency insights. Similarly, Polynesia \cite{boroumand2021polynesia} introduces an in-memory HTAP system that partitions transactional and analytical workloads into separate "islands," maintaining dedicated data replicas for each to optimize performance.

In comparison to most above research on data transformations, our research on TE-LSM and Mycelium is not about improvements of transformation techniques themselves, but rather applying them at a different time at minimal cost or when that cost can be (at least partially) hidden.

\textbf{Piggyback data transformations on compaction}: To the best of our knowledge, the work by Saxena et al. on the Real-Time LSM-Tree is the only research that leverages the compaction process to piggyback data transformations, specifically transitioning data from a row-wise to a columnar format. The Real-Time LSM-Tree is a data lifecycle-aware storage engine that enables varying data layouts across LSM-tree levels, progressively shifting from purely row-oriented to purely column-oriented storage. While Saxena's work has inspired our exploration of data transformations targeting the physical design of data, our research generalizes this approach to a much broader variety of transformations.

\section{Conclusion}
In this paper, we presented Mycelium, a novel approach that hides the cost of data transformations by embedding them within the compaction process of Log-Structured Merge-trees (LSM-trees). Unlike traditional methods focused on minimizing compaction costs, Mycelium improves overall system performance by offloading additional data transformation tasks to the compaction phase. This design enables inherently write-optimized data stores like LSM-trees to become more read-optimized, significantly enhancing performance for future read operations.

Our experimental evaluation demonstrates that Mycelium delivers substantial improvements in read latency, especially for analytic (column-accessing) queries, while introducing only minimal overhead to write throughput. Furthermore, Mycelium substantially outperforms naive approaches that handle data transformations outside the compaction process, as evidenced by the much lessened write throughput impact.

The seamless integration of data transformation into the compaction process in Mycelium creates opportunities for implementing more sophisticated transformation logic within storage systems. This flexible design paves the way for further optimizations in balancing transformation complexity with overall system performance 

Future work will focus on extending Mycelium's capabilities to support a broader range of transformations, particularly in ETL workflows, adaptive indexing mechanisms, dynamic column group splitting, and integration with distributed storage systems. Additionally, investigating the relationship between transformation complexity and compaction strategies could yield deeper insights into achieving optimal performance across diverse workloads.

In conclusion, Mycelium represents a significant advancement in embedding in-place data transformation with storage engine operations, offering an efficient and scalable solution for modern data-intensive applications.
\section*{Availability}
Source code is available but hidden due to anonymity.

\bibliographystyle{plain}
\bibliography{\jobname}

\appendix

\section{Algorithms}
\label{appendix:algo}
In this section, we show the core algorithms used in Mycelium to integrate various data transformations within compaction process.

Algorithm \ref{alg:link_cf_to_transformers} creates internal destination column families, and link those column families to the transformers to initiate seamless transformation-embedded compactions.
\begin{algorithm}[H]
\caption{\textsc{LinkTransformers}}
\label{alg:link_cf_to_transformers}
\begin{tabular}{@{}p{1.2cm}p{7cm}@{}}
\textbf{Input:} & src\_cf - User-facing column family  \\
                & xformers - List of transformers \\
\textbf{Output:} & status - Indicating Success or Failure \\
\end{tabular}
\begin{algorithmic}[1]
\STATE \texttt{xformerSortedList $\leftarrow$ ValidateAndSort(xformers)}
\STATE \texttt{colFamQ $\leftarrow$ Queue(src\_cf)}
\WHILE{\texttt{colFamQ is not empty}}
    \STATE \texttt{colFamQSize $\leftarrow$ colFamQ.size()}
    \STATE \texttt{xformer $\leftarrow$ xformerSortedList[0]}
    \STATE \texttt{xformerSortedList $\leftarrow$ xformerSortedList[1:]}
    \FOR{\texttt{i $\leftarrow$ 0 \textbf{to} colFamQSize - 1}}
        \STATE \texttt{curr $\leftarrow$ colFamQ.front()}
        \STATE \texttt{LinkTransformer(curr, xformer)}
        \STATE \texttt{destCFs $\leftarrow$ CreateInternalCFs(curr, xformer)}
        \IF{xformerSortedList.size() $>$ 0}
            \STATE \texttt{colFamQ.enqueue(destCFs)}
        \ENDIF
        \STATE \texttt{colFamQ.pop()} \quad \COMMENT{remove the processed column family}
    \ENDFOR
\ENDWHILE
\STATE \textbf{return} \texttt{Success}
\end{algorithmic}
\end{algorithm}

Algorithm \ref{alg:performing_transformations} presents a snippet of applying transformations within a compaction job. For column families without an associated transformer, the process is a no-op; otherwise, the specified transformations are applied, and the transformed outputs are returned.
\begin{algorithm}[H]
\caption{\textsc{ApplyTransformations}}
\label{alg:performing_transformations}
\begin{tabular}{@{}p{1.2cm}p{5.5cm}@{}}
\textbf{Input:} & subcompaction - The current compaction segment, containing an iterator over data and the associated column family information. \\
\textbf{Output:} & outputs - A list of outputs after applying the transformer to the compaction data. \\
\end{tabular}
\begin{algorithmic}[1]
\STATE \texttt{outputs $\leftarrow$ []} \quad \COMMENT{Initialize the list of\\ \qquad \qquad \quad \quad \quad \quad transformed outputs}
\STATE \texttt{cfd $\leftarrow$ subcompaction.column\_family}
\IF{\texttt{cfd.transformer $\ne$ nullptr}}
    \STATE \texttt{outputs $\leftarrow$ transformer.Transform(} \\
           \qquad \qquad \qquad \texttt{subcompaction.iter, cfd.schemaInfo)} 
\ENDIF
\STATE \textbf{return} \texttt{outputs} \qquad \COMMENT{Return the list of\\ \qquad \qquad \quad \quad \quad \quad transformed outputs}
\end{algorithmic}
\end{algorithm}

Algorithm \ref{alg:install-outputs} demonstrates the process of installing the results of a transformation-embedded compaction. When the column family being compacted has an associated transformer, the transformed outputs are written to SST files in the internal destination column families. Meanwhile, the input files that were compacted are deleted and their metadata is removed from the source column family.
\begin{algorithm}[H]
\caption{\textsc{InstallTransformedOutputs}}
\label{alg:install-outputs}
\begin{tabular}{@{}p{1.2cm}p{6.5cm}@{}}
\textbf{Input:} & subcompaction - The current compaction segment, containing the transformed outputs and the compacted input files. \\
\textbf{Output:} & status - Indicating Success or Failure \\
\end{tabular}
\begin{algorithmic}[1]
\STATE \texttt{cfd $\leftarrow$ subcompaction.column\_family}
\IF{\texttt{xformer $\ne$ nullptr}}
    \STATE \texttt{destCFs $\leftarrow$ GetDestinationColumnFamilies(cfd)}
    \STATE \texttt{InstallOutputs(subcompaction.outputs, destCFs)}
    \STATE \texttt{DeleteInputs(subcompaction.inputs, cfd)}
\ENDIF
\STATE \textbf{return} \texttt{Success}
\end{algorithmic}
\end{algorithm}

\section{Cost Model}
\label{appendix:cost-model}
In this section, we present the cost model analysis. We run the analysis with examples in each subsection. 

Before we start the analysis, we use Table \ref{tab:symbols_table} to list the symbols used in the analysis. We assume that data arrives in random key ranges, causing all sorted runs in Level 0 to overlap.

\begin{table}[h!]
    \centering
    \resizebox{\columnwidth}{!}{  
        \begin{tabular}{|c|l|}
        \hline
        N   & Total size of data \\ \hline
        B   & Size of write buffer \\ \hline
        T   & Size factor between adjacent levels in the LSM-tree \\ \hline
        R   & Size of a data record \\ \hline
        blksz   & Disk block size \\ \hline
        Z   & Number of sst files in Level-0 \\ \hline
        $P_{false}$ & false positive reporting probability of bloom filter \\ \hline
        L   & Number of levels in the LSM-tree \\ \hline
        \end{tabular}
    }
    \caption{Symbols used in the cost analysis}
    \label{tab:symbols_table}
\end{table}

\subsubsection{Write Throughput} Let $WB_{disk}$ represent the disk write bandwidth, $RB_{disk}$ the disk read bandwidth, and $T_r$ the data transformation throughput (size of data processed per second). Let $WA_{CWT}$ denote the write amplification for compaction without transformations (CWT), and $WA_{TEC}$ the write amplification for transformation-embedded compaction (TEC). In the case of CWT, assuming that in the case of CWT, the $WB_{disk}$ is less than $WR_{disk}$, the maximum write throughput can be expressed as: 

\begin{equation}
    W_{\text{max,CWT}} = \frac{WB_{disk}}{WA_{CWT}} = \frac{WB_{disk}}{1+\frac{T}{T-1}\log_T \left(\frac{N}{B} \right)}
\end{equation}

For TEC, the maximum write throughput is given by:
\begin{equation}
    W_{\text{max,TEC}} = \frac{\min \left( WB_{\text{disk}}, \frac{RB_{\text{disk}} \cdot T_r}{RB_{\text{disk}} + T_r} \right)}{WA_{\text{TEC}}}
\end{equation}

Where:  
\[
WA_{\text{TEC}} = WA_{\text{CWT}} + n \quad (1 \leq n < T/2)
\]
The additional write amplification \( n \) accounts for extra writes incurred during cross-column family compaction.  

Assuming the disk write bandwidth is determined by the minimum of \( WB_{\text{disk}} \) and \( \frac{RB_{\text{disk}} \cdot T_r}{RB_{\text{disk}} + T_r} \), the maximum write throughput for TEC simplifies to:
\begin{equation}
    W_{\text{max,TEC}} = \frac{WB_{\text{disk}}}{1 + n + \frac{T}{T-1}\log_T \left(\frac{N}{B} \right)}
\end{equation}

To demonstrate the application of the above equations, consider the following example:  

\begin{itemize}[itemsep=0pt, topsep=0pt, left=0pt]
    \item Total data size \( N = 100 \, \text{TB} \)  
    \item Write buffer size \( B = 64 \, \text{MB} \)  
    \item Size ratio \( T = 10 \)  
    \item Extra writes during cross-column family compaction \( n = 2 \)  
    \item SSD write bandwidth \( WB_{\text{disk}} = 417 \, \text{MB/s} \)  
\end{itemize}

The maximum write throughput for CWT is calculated as:  
\[
W_{\text{max,CWT}} = \frac{417}{1 + \frac{10}{9} \log_{10} \left(\frac{100 \, \text{TB}}{64 \, \text{MB}} \right)} \approx 52.75 \, \text{MB/s}
\]

For TEC, the maximum write throughput is:  
\[
W_{\text{max,TEC}} = \frac{417}{1 + 2 + \frac{10}{9} \log_{10} \left(\frac{100 \, \text{TB}}{64 \, \text{MB}} \right)} \approx 42.10 \, \text{MB/s}
\]

This results in an approximate \textbf{20\% reduction} in write throughput from CWT to TEC.

\subsubsection{Point Query}  
We analyze the cost of point queries in two scenarios:  
\begin{enumerate}[itemsep=0pt, topsep=0pt]
    \item When the entire value associated with the key is required.  
    \item When only a single field is needed.  
\end{enumerate}  

We denote the former as PQRA and the latter as PQRC.  

Let \( L \) represent the number of levels in the LSM tree, determined by:  
\[
L = \log_{T}\left(\frac{N}{B}\right)
\]  
where \( N \) is the total data size, \( B \) is the write buffer size, and \( T \) is the size ratio.  

Let \( s_i \) represent the number of internal column families serving as compaction destinations, and \( n \) the number of cross-column family compactions the data undergoes. The cost of point query retrieving the entire row can be expressed as:

\[
C_{\text{PQRA}} = (L + Z \cdot \left(1 + n \right)) \cdot P_{\text{false}} + \frac{R}{\text{blksz}} \cdot s_n  \quad (1 \leq n < T/2)
\]

And the cost of point query accessing only a single field is given as:

\[
C_{\text{PQRC}} = (L + Z \cdot \left(1 + n \right)) \cdot P_{\text{false}} + \frac{R}{\text{blksz}}
\]

To further illustrate the above equation, we provide two examples:  

\begin{enumerate}[itemsep=0pt, topsep=0pt]
    \item \textbf{TEC with \textbf{convert} transformations:}  
    \begin{itemize}[itemsep=0pt, topsep=0pt, left=0pt]
        \item \( L = 6 \), \( Z = 2 \), \( n = 1 \), \( s_i = 1 \), \( P_{\text{false}} = 0.01 \), \( R = 5000 \, \text{bytes} \), Block size \( \text{blksz} = 4096 \, \text{bytes} \).
        \item After conversion, \( R \) is reduced by 30\%.
        \item Substituting these values yields the cost for both PQRA and PQRC for TEC with embedded \textbf{convert} transformations as approximately 1.1 \text{block reads}.
    \end{itemize}
    
    \item \textbf{TEC with \textbf{split} transformations:}  
    \begin{itemize}[itemsep=0pt, topsep=0pt, left=0pt]
        \item \( L = 5 \), \( Z = 2 \), \( n = 3 \), \( s_n = 8 \), \( P_{\text{false}} = 0.01 \), \( R = 5000 \, \text{bytes} \), Block size \( \text{blksz} = 4096 \, \text{bytes} \).
        \item Substituting the values yields the worst-case PQRA and PQRC costs for TEC with embedded \textbf{split} transformations, respectively, as 8.13 \text{block reads} and 1.13 \text{block reads}.
    \end{itemize}
\end{enumerate}

For comparison, in the case of CWT, this cost is approximately \( 2.08 \, \text{block reads} \).

\subsubsection{Range Query}  
We assume that most range queries operate on a projection of fields. Thus, we analyze the cost of range queries under this assumption.  

Let \( m \) represent the selectivity, defined as the number of unique entries that fall within the target key range.  

- For CWT, there are \( \frac{m}{T^{L-i}} \) entries scanned at each level of the LSM-tree.  
- For TEC, the total number of entries scanned is:  
\[
\frac{m \cdot (n + T^i)}{T^L} \quad (1 \leq n < T/2)
\]  

The corresponding number of block reads for CWT is given by:  
\[
C_{\text{RQ,CWT}} = \frac{m \cdot R}{\text{blksz}} \cdot \sum_{i=0}^{L} T^{i-L}
\]  

For TEC, the block read cost is:  
\[
C_{\text{RQ,TEC}} = \frac{m}{\text{blksz}} \cdot \left(\frac{\sum_{j=1}^{n} R_{j}}{T^L} + R_{n} \cdot \sum_{i=0}^{L} T^{i-L} \right)
\]  

Using the values from the previous point query examples and setting the selectivity \( m = 100 \), the range query cost for TECs applying the \textbf{convert} and \textbf{split} transformations is calculated as:  

\begin{enumerate}[itemsep=0pt, topsep=0pt]
    \item \textbf{TEC embedded with convert transformation:} \(\approx 97.78\) block reads
    \item \textbf{TEC embedded with split transformation:} \(\approx 17.78\) block reads
\end{enumerate}  

For comparison, the range query cost for CWT is approximately \( 138.88 \) block reads, resulting in:  
- An \( 87.2\% \) improvement for the \textbf{split} transformation case  
- A \( 29.6\% \) improvement for the \textbf{convert} transformation case 

\subsubsection{Space Amplification} Lastly, we analyze the additional space amplification introduced by transformation-embedded LSM-trees compared to LSM-trees that perform no transformations.  

In the worst-case scenario for an LSM-tree without transformations (CWT), maximum space amplification occurs when every record in level \( i-1 \) invalidates a record in level \( i \) during merging. This results in a space amplification of \( O(1/T) \).  

Let the size of a value in a key-value pair be denoted by \( V \), and the size of a key by \( K \). The following analysis examines three transformation scenarios that impact space amplification:

\begin{itemize}[itemsep=0pt, topsep=0pt, left=0pt]
    \item \textbf{splitting column groups}: When column groups are split, each resulting group stores the key separately, leading to the key being written multiple times.  This results in additional space amplification for a transformation-embedded LSM-tree compared to an LSM-tree without transformations. The additional space amplification is given by the following equation, where \( s_n \) represents the number of split column families at the last cross-column family compaction level:  

    \[
        SPAmp_{\text{split}} = O\left(\frac{K \cdot (s_n - 1) \cdot N}{R \cdot T}\right)
    \]  

    \item \textbf{converting data format}: In the case of data format conversion, the additional space amplification could be greater than (space is more amplified) or less than  (less space is used) 1 depending on the final size of data after the conversion. Let \texttt{\(R'\)} denote the size of data after the transformation, and \texttt{\(R\)}) the original size of data, the space amplification is given by: 
    \[
       SPAmp_{\text{convert}} =  O(\frac{N \cdot R'}{R \cdot T}) 
    \]
    \item \textbf{creating index}: Secondary indexes are not considered as contributing to the space amplification of the primary data. Therefore, the space amplification for creating an index in a transformation-embedded LSM-tree remains the same as that of an LSM-tree without transformations: \texttt{\(O(\frac{1}{T})\)}.
\end{itemize}

\section{Query Performance of Mycelium-Identity and Mycelium-Augment\\}
\label{appendix:query-perf-M-Iden-Aug}
We run the query latency tests (Q2, Q3, Q6, and Q7) on both Mycelium-Identity and Mycelium-Augment. Since Mycelium-Identity performs a no-op transformation, its query performance is expected to closely match that of the RocksDB baseline. Similarly, for Mycelium-Augment, because these queries operate solely on the primary data, their performance should also resemble the RocksDB baseline (Figure \ref{fig:noop-baw}). The slight increase in latency is due to the additional transformed column families created to support transformation-embedded compactions. However, this overhead is minimal, resulting in only marginally higher latencies.

\begin{figure}[H]
    \centering
    \includegraphics[width=\linewidth]{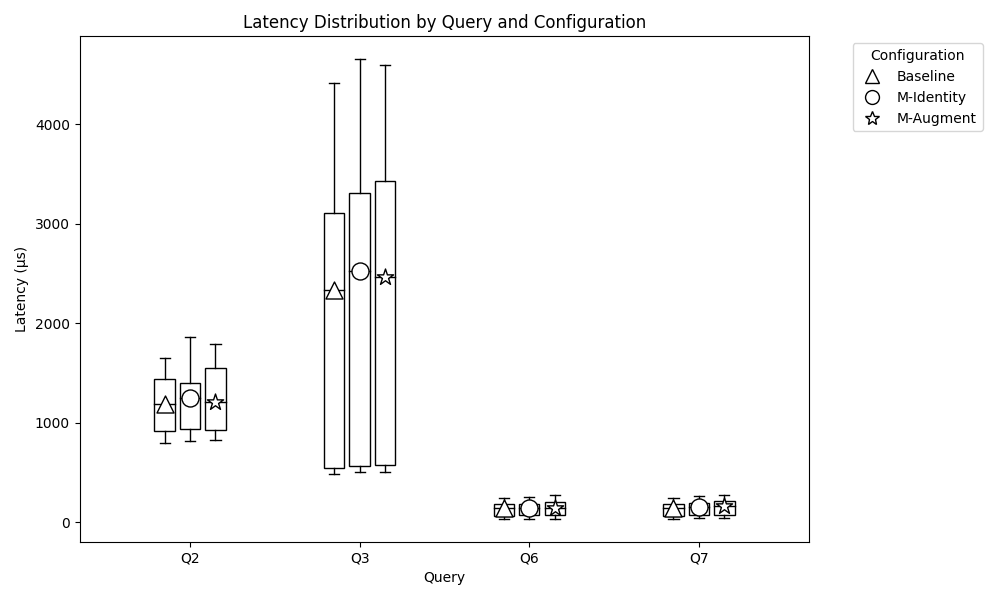}
    \caption{Mycelium-Identity (M-Identity) and Mycelium-Augment (M-Augment) have closely-aligned performance compared to the RocksDB baseline for Q2, Q3, Q6, and Q7}
    \label{fig:noop-baw}
\end{figure}

\section{RocksDB/Mycelium configuration parameters}
\label{appendix:config}
In this section, we share the configuration parameters we use for the evaluation in the following:

\lstset{
    language=C++,
    basicstyle=\ttfamily\small,           
    keywordstyle=\color{blue}\bfseries,   
    commentstyle=\color{gray}\itshape,    
    stringstyle=\color{red},              
    numbers=left,                         
    numberstyle=\tiny\color{gray},        
    stepnumber=1,                         
    numbersep=5pt,                        
    showstringspaces=false,               
    tabsize=4,                            
    breaklines=true,                      
    breakatwhitespace=false,              
    frame=single,                         
    columns=fullflexible                  
}

\begin{lstlisting}
options_.create_if_missing = true;
options_.enable_pipelined_write = true;
options_.max_open_files = -1;
options_.env->SetBackgroundThreads(16, rocksdb::Env::Priority::LOW);
options_.env->SetBackgroundThreads(8, rocksdb::Env::Priority::HIGH);
options_.max_background_compactions = 16;
options_.max_background_flushes = 8;
options_.max_subcompactions = 16;
options_.compaction_style = rocksdb::kCompactionStyleLevel;
options_.SetTransformerType(rocksdb::TransformerType::CONVERTER);
options_.SetInputOutputDataType(inputDataType, outputDataType);
options_.write_buffer_size = 128 * 1024 * 1024;
options_.max_write_buffer_number = 8;
options_.max_bytes_for_level_base = 256 * 1024 * 1024;
options_.target_file_size_base = 256 * 1024 * 1024;
options_.level0_file_num_compaction_trigger = 4;
options_.level0_slowdown_writes_trigger = 30;
options_.level0_stop_writes_trigger = 64;

rocksdb::BlockBasedTableOptions table_options;
table_options.block_cache = rocksdb::NewLRUCache(512 * 1024 * 1024);
table_options.filter_policy.reset(rocksdb::NewBloomFilterPolicy(10, false));
options_.table_factory = std::shared_ptr<rocksdb::TableFactory>(
    rocksdb::NewBlockBasedTableFactory(table_options)
);
\end{lstlisting}

\end{document}